\documentclass[11pt]{article}
\usepackage{graphicx}

\newcommand{\De}{\Delta}

\newcommand{\bn}{{\bf n}}

\newcommand{\be}{\begin{equation}}
\newcommand{\ee}{\end{equation}}

\begin{document}

\title{Cosmological constraints from Microwave Background Anisotropy
and Polarization}

\author{Alessandro Melchiorri\\
 Universita' di Roma ``La Sapienza''\\ 
Ple Aldo Moro 2, 00185, Rome, Italy}

\maketitle

\begin{abstract}

The recent high-quality measurements of the Cosmic Microwave Background 
anisotropies and polarization provided 
by ground-based, balloon-borne and satellite experiments have presented 
cosmologists with the possibility of studying the large scale properties of 
our universe with unprecedented precision.
Here I review the current status of observations and constraints
on theoretical models.

\end{abstract}

\section{Introduction}\label{s1}

The nature of cosmology as a mature and testable science
lies in the realm of observations of Cosmic Microwave Background 
(CMB) anisotropy and polarization. The recent high-quality 
measurements of the CMB anisotropies provided by ground-based, 
balloon-borne  and satellite experiments have indeed presented cosmologists 
with the possibility of studying the large scale properties of our 
universe with unprecedented precision.\\ 
An increasingly complete cosmological image arises as the key parameters 
of the cosmological model  have now been constrained within a few percent 
accuracy. The impact of these results in different sectors than
cosmology has been extremely relevant since CMB studies can 
set stringent constraints on the early thermal history of the universe 
and its particle content.
For example, important constraints have been placed in fields 
related to particle physics or quantum gravity 
like neutrino physics, extra dimensions, and super-symmetry theories.\\
In the next couple of years, new and current on-going experiments
will provide datasets with even higher quality and information.
In particular, accurate measurements of the CMB polarization statistical 
properties represent a new research area. 
The CMB polarization has been detected by two experiments, but 
remains to be thoroughly investigated. In conjunction with our extensive 
knowledge about the CMB temperature anisotropies, new constraints on the 
physics of the early universe (gravity waves, isocurvature perturbations, 
variations in fundamental constants) as well as late universe phenomena 
(reionization, formation of the first objects, galactic foregrounds) will be 
investigated with implications for different fields ranging from particle 
physics to astronomy.\\
Moreover, new CMB observations at small (arcminute) angular scales will 
probe secondary fluctuations associated  with the first nonlinear objects.
This is where the first galaxies and the
first quasars may leave distinct imprints in the CMB and where an interface
between cosmology and the local universe can be established. 
In this proceedings I will briefly review the current status of CMB 
observations, I discuss the agreement with the current theoretical
scenario and I will finally draw some conclusions.

\section{The standard picture.}

The standard model of structure formation (described in great
detail in several reviews, see e.g. \cite{review}, 
\cite{review2}, \cite{review3}, \cite{review4}, \cite{review5}).
relies on the presence of a background of tiny 
(of order $10^{-6}$) primordial density 
perturbations on all scales (including those larger than
the causal horizon). \\
This primordial background of perturbations is assumed gaussian, adiabatic, 
and nearly scale-invariant as generally predicted by the inflationary 
paradigm. Once inflation is over, the evolution of all Fourier mode 
density perturbations is linear and passive (see \cite{review5}).\\
Moreover, prior to recombination, a given Fourier mode begins
oscillating as an acoustic wave once the horizon overtakes its wavelength.
Since all modes with a given wavelength begin evolving simultaneously
the resulting acoustic oscillations are phase-coherent, leading to
a structure of peaks in the temperature and polarization power spectra of
the Cosmic Microwave Background (\cite{Peeb1970}, \cite{SZ70}, 
\cite{wilson}).\\
The anisotropy with respect to the mean temperature $\Delta T=T-T_0$
 of the CMB sky in the direction $\bn$ measured at
time $t$ and from the position $\vec x$ can be expanded in 
spherical harmonics:

\be
{\De T\over T_0}(\bn,t,\vec x) = \sum_{\ell=2}^\infty\sum_{m=-\ell}^{m=\ell}
	 a_{\ell m}(t, \vec x) Y_{\ell m}(\bn)~,
\ee

If the fluctuations are Gaussian all the statistical information is
contained in the $2$-point correlation function. In the case of isotropic
fluctuations, this can be written as:

\be
\left \langle{\De T\over T_0}({\bn_1}){\De T\over T_0}({\bn_2})\right\rangle =
 {1\over 4\pi} \sum_\ell (2\ell+1)C_\ell P_\ell(\bn_1\cdot\bn_2)~.
\ee

where the average is an average over "all the possible universes"
i.e., by the ergodic theorem, over $\vec x$. The CMB power
spectrum $C_\ell$ are the ensemble average of the coefficients
$a_{\ell m}$,
\[ C_\ell = \langle|a_{\ell m}|^2\rangle ~. \]

A similar approach can be used for the cosmic microwave background
polarization and the cross temperature-polarization correlation
functions.
Since it is impossible to measure ${\De T\over T_0}$ in every position in
the universe, we cannot do an ensemble average.
This introduces a fundamental limitation for the precision of a 
measurement (the cosmic variance) which is 
important especially for low multipoles. 
If the temperature fluctuations are Gaussian, the $C_{\ell}$ have 
a chi-square distribution with $2\ell+1$ degrees of freedom and 
the observed mean deviates from the ensemble average by 

\be 
{{\Delta C_\ell} \over C_\ell} = 
	\sqrt{2\over 2\ell + 1}~.  \label{2cv}
\ee

Moreover, in a real experiment, one never obtain complete sky 
coverage because of the limited amount of observational time 
(ground based and balloon borne
experiments) or because of galaxy foreground contamination 
(satellite experiments). 
All the telescopes also have to deal with the noise of the detectors 
and are obviously not sensitive to scales smaller than the 
angular resolution.\\

\begin{figure}
\centerline{\includegraphics[width=4.0in]{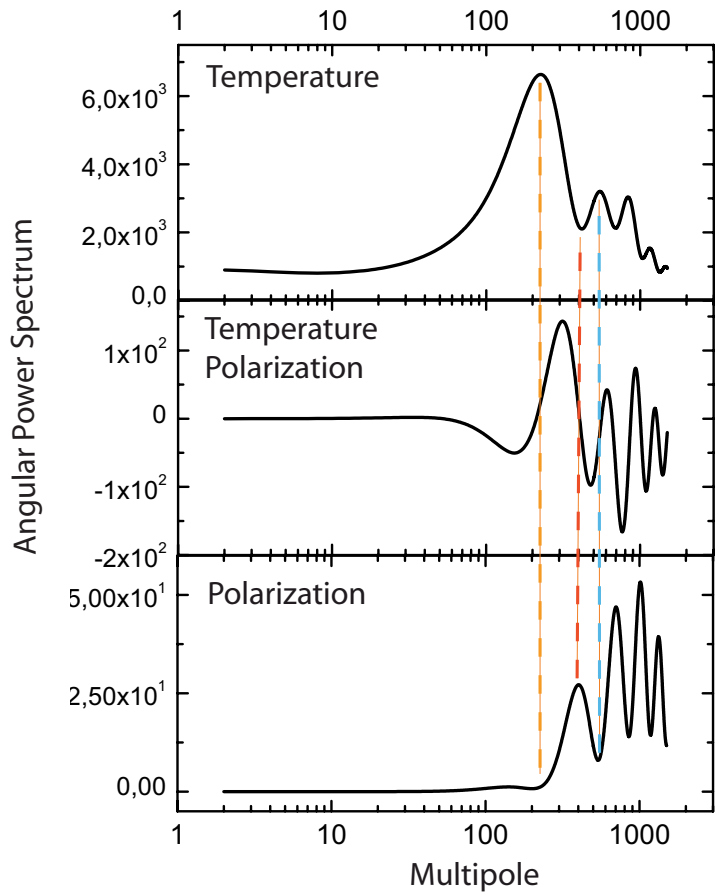}}
\caption{\label{fig1} Theoretical predictions for the CMB
temperature, polarization and cross temperature-polarization power
spectra in the case of the standard model of structure formation.
The peaks in the temperature and polarization spectra are alternate.}
\end{figure}

\begin{figure}
\centerline{\includegraphics[width=4.0in]{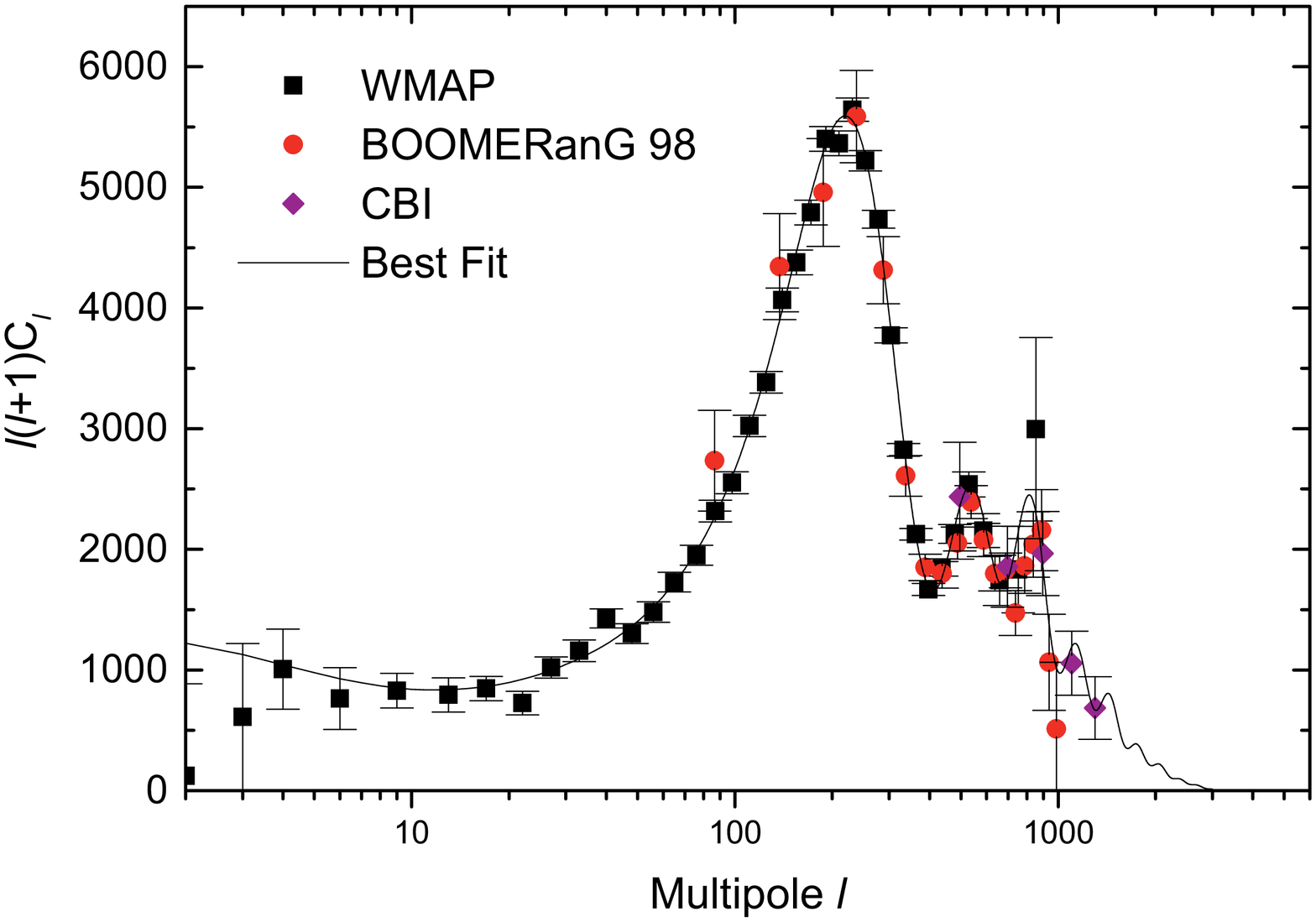}}
\caption{\label{fig2} Cosmic Microwave Anisotropies data
from WMAP, BOOMERanG and CBI experiments and the theoretical
best fit. There is an excellent agreement between data and theory.}
\end{figure}

\noindent In Figure $1$ we plot the theoretical prediction for CMB 
temperature and polarization power spectra and the 
cross-correlation between temperature and polarization in the case of the 
so-called 'concordance' model. The power spectra show an unique structure: 
The temperature power spectrum is flat on large scale
while shows oscillations on smaller scales. The polarization and
cross temperature-polarization spectra are also showing oscillations
on smaller scales, but the signal at large scale is expected to be 
negligible.\\
Different mechanisms are responsible
for the oscillations on small scale. 
On the scale of the thickness of the last scattering
surface, temperature anisotropies are more coupled to density and
gravity perturbations, while polarization is more coupled to velocity
perturbations. Gravity and density perturbations obey a cosine 
function, while velocity perturbations follow a sine function. 
The most striking observational prediction of this is
the out-of-phase position of the peaks and dips in the temperature
and polarization power spectra. The peaks and dips of the
cross-correlation spectra fall in the middle (see e.g.
\cite{koso2}).\\
The shape of the power spectra depends on the value
of the cosmological parameters assumed in the theoretical
computation. Since the overall picture must be consistent, 
the cosmological parameters determined indirectly from CMB observations 
must agree with the values inferred from independent observations (Big Bang
Nucleosynthesis, Galaxy Surveys, Ly-$\alpha$ forest clouds, simulations,
etc etc.).

\section{The latest measurements.}

The last years have been an exciting period for
the field of the CMB research. 
With the TOCO$-97/98$ (\cite{torbet},\cite{miller}) 
and BOOMERanG-$97$ (\cite{mauskopf}) experiments a firm detection of
a first peak in the CMB angular power spectrum 
on about degree scales has been obtained. 
In the framework of adiabatic Cold Dark Matter (CDM) models, the
position, amplitude and width of this peak provide strong supporting 
evidence for the inflationary predictions of
a low curvature (flat) universe and a scale-invariant primordial 
spectrum (\cite{knox}, \cite{melchiorri}, \cite{tegb97}).\\
The subsequent data from BOOMERanG LDB (\cite{netterfield}), 
DASI (\cite{halverson}), MAXIMA (\cite{lee}), 
VSAE (\cite{grainge}) and Archeops 
(\cite{benoit}) have provided further evidence for the presence of the 
first peak and refined the data at larger multipole hinting
towards the presence of multiple peaks in the spectrum.
Moreover, the very small scale observations made by the
CBI (\cite{pearson}) and ACBAR (\cite{acbar}) experiments 
have confirmed the presence of a damping tail, while the
new DASI results presented the first evidence for polarization
(\cite{dasipol}).\\
The combined data clearly confirmed the model
prediction of acoustic oscillations in the primeval plasma 
and shed new light on several cosmological and 
inflationary parameters ( see e.g. 
\cite{anze}, \cite{odman}, \cite{wang2}).\\
The recent results from the WMAP satellite experiment (see Figure $2$)
(\cite{bennett}) have confirmed in a spectacular way all these 
previous results with a considerable reduction of the error
bars. In particular, the amplitude and position of the
first two peaks in the spectrum are now determined with a precision 
about $6$ times better than before (\cite{page}).\\
Furthermore, the WMAP team released the first high quality measurements
of the temperature-polarization spectrum \cite{kogut}.
In particular. WMAP detected large angular scale polarization, 
indicative of an earlier reionization of the intergalactic medium 
(\cite{kogut}). 
The presence of polarization at intermediate angular scale helps
in discriminating inflationary models from causal scaling seed 
toy models (see e.g.\cite{peiris}). Moreover, the  position of the first 
anti-peak and second peak in the spectrum are also in agreement with the
prediction of inflation (\cite{dode}).\\
As main intriguing discrepancy, the WMAP data shows (in agreement with the 
previous COBE data) a lower temperature quadrupole than expected.
The statistical significance of this discrepancy is still unclear
(see e.g. \cite{costa}, \cite{efstathiou2}, \cite{dore}).\\
The CMB anisotropies measured by WMAP are also in good agreement 
with the standard inflationary prediction of gaussianity 
(\cite{komatsu}).

\section{CMB constraints on the standard model.}

In principle, the standard scenario of structure formation based on adiabatic
primordial fluctuations can depend on more than $11$ parameters.\\
However for a first analysis and under the assumption of a flat universe
which is already well consistent with CMB data, it is 
possible to restrict ourselves to just $5$ parameters: the tilt of primordial 
spectrum of scalar perturbations $n_S$, the optical depth of the universe 
$\tau_c$, the physical energy densities in baryons and dark matter
$\omega_b=\Omega_bh^2$ and $\omega_{dm}=\Omega_{dm}h^2$ and 
the Hubble parameter $h$.

\begin{table}
\begin{tabular}{lllll}
\hline
& WMAP & WMAPext &  WMAPext+LSS  & Pre-WMAP+LSS \\
\hline
$\Omega_b h^2$ &\ensuremath{0.024 \pm 0.001} &
\ensuremath{0.023 \pm 0.001}&
\ensuremath{0.023 \pm 0.001}&
\ensuremath{0.021 \pm 0.003} \\
\hline
$\Omega_m h^2$ &\ensuremath{0.14 \pm 0.02} &
\ensuremath{0.13 \pm 0.01}&
\ensuremath{0.134 \pm 0.006}&
\ensuremath{0.14 \pm 0.02} \\
\hline
$h$ &\ensuremath{0.72 \pm 0.05} &
\ensuremath{0.73 \pm 0.05}&
\ensuremath{0.73 \pm 0.03}&
\ensuremath{0.69 \pm 0.07} \\
\hline
$n_s$ &\ensuremath{0.99 \pm 0.04} &
\ensuremath{0.97 \pm 0.03}&
\ensuremath{0.97 \pm 0.03}&
\ensuremath{0.97 \pm 0.04} \\
\hline
$\tau_c$ &\ensuremath{0.166^{+ 0.076}_{- 0.071}} &
\ensuremath{0.143^{+ 0.071}_{- 0.062}}&
\ensuremath{0.148^{+ 0.073}_{- 0.071}}&
\ensuremath{0.07^{+ 0.07}_{- 0.05}} \\
\hline
\end{tabular}
\caption{Current constraints on the $5$ parameters of the standard model
(flat universe). The WMAP results are taken from Spergel et al. 2003, 
the previous results are taken from Melchiorri and Odman 2003 (see also
Slosar et al. 2003 and Wang et al. 2003).}
\end{table}

In Table $1$ we report the constraints on these parameters 
obtained by the WMAP team (see \cite{spergel} and \cite{verde})
in $3$ cases: WMAP only,
WMAP+CBI+ACBAR and WMAP+CBI+ACBAR+LSS. Also, for comparison,
we present the CMB+LSS results previous to WMAP in the forth column.\\
As we can see a value for the baryon density $\omega_b =0.020\pm0.002$ 
as predicted by Standard Big Bang Nucleosynthesis (see e.g.\cite{cyburt})
is in very good agreement with the WMAP results.
The WMAP data is in agreement with the previous results 
and its inclusion reduces the error bar on this parameter by 
a factor $3$.\\
The amount of cold dark matter is also well constrained by the CMB data. 
The presence of power around the third peak is crucial in this sense, 
since it cannot be easily accommodated in models based on just baryonic 
matter (see e.g. \cite{melksilk} and references therein). 
As we can see, including the CMB data on those scales (not sampled
by WMAP) halves the error bars.
WMAP is again in agreement with the previous determination 
and its inclusion reduces the error bar on this parameter by 
of  factor $3$-$4$.\\
These values implies the existence of a cosmological constant
at high significance with $\Omega_{\Lambda}=0$, $\Omega_{M}=1$
excluded at $5$-$\sigma$ from the WMAP data alone and at
$\sim 15$-$\sigma$ when combined with supernovae data.
A cosmological constant is also suggested from the evidence of 
correlations of the WMAP data with large scale structure 
(\cite{scranton}, \cite{nolta}, \cite{boughn},
\cite{afshordi}).\\
Under the assumption of flatness, is possible to constrain 
the value of the Hubble parameter $h$. The constraints on
this parameter are in very well agreement with the
HST constraint (\cite{freedman}).\\
An increase in the optical depth $\tau_c$ after recombination 
by reionization damps the amplitude of the CMB peaks. 
This small scale damping can be somewhat compensated by an
increase in the spectral index $n_S$. This leads to a nearly perfect
degeneracy between $n_S$ and $\tau_c$ and, in practice, no 
significant upper bound on these parameters can be placed from temperature 
data. However, large scale polarization data, as measured by WMAP, and
LSS data can break this degeneracy. At the same time, inclusion of
galaxy clustering data can determine $n_S$ and further break the
degeneracy. As we can see, the current constraint on the spectral
index is close to scale invariance ($n_S \sim 1$) as predicted
by inflation. The best-fit value of the optical depth determined by WMAP
is slightly higher but consistent in between $1-\sigma$ with supercomputer 
simulations of reionization processes ($\tau_c \sim 0.10$,
see e.g. \cite{ciardi}).

\section{Constraints on possible extensions of the standard model.}

The standard model provides a reasonable fit to the data.
However it is possible to consider several modifications characterized
by the inclusion of new parameters. The data considered here 
doesn't show any definite evidence for those modifications, providing more
a set of useful constraints. The present status can be briefly
summarized as follows:

\begin{itemize}

\item Running of the Spectral Index

The possibility of a scale dependence of the
scalar spectral index, $n_S(k)$, has been considered
in various works (see e.g. \cite{kosowsky}, \cite{copeland}, 
\cite{lythcovi}, \cite{doste}). Even though this dependence is considered to 
have small effects on CMB scales in most of the slow-roll inflationary models, 
it is worthwhile to see if any useful constraint can be obtained.
The present CMB data is at the moment compatible with no
scale dependence (\cite{kkmr},\cite{bridle}), however, joint analyses with 
other datasets (like Lyman-$\alpha$) shows a $\sim 2-\sigma$ evidence for a 
negative running (\cite{peiris}). At the moment, the biggest case
against running comes from reionization models, which are unable
to reach the large optical depth observed in this case
by WMAP ($\tau \sim 0.17$, see \cite{spergel}) with the suppressed power 
on small scales (see e.g. \cite{cen}).

\item Gravitational Waves.

The metric perturbations created during inflation belong to two types:
{\it scalar} perturbations, which couple to the stress-energy of 
matter in the universe and form the ``seeds'' for structure formation 
and {\it tensor} perturbations, also known as 
gravitational wave perturbations.
A sizable background of gravity waves 
is expected in most of the inflationary scenarios and
a detection of the GW background 
can provide information on the second derivative
of the inflaton potential and shed light on the physics at
$\sim 10^{16} Gev$.

The amplitude of the GW background is weakly constrained
by the current CMB data. However, when information from BBN, local
cluster abundance and galaxy clustering are included, an upper limit
of about $r = C_2^T/C_2^S < 0.5$ (no running) is obtained
(see e.g. \cite{spergel}, \cite{kkmr}, \cite{barger}, \cite{leach}).

\item Isocurvature Perturbations
 
Another key assumption of the standard model is that the primordial 
fluctuations were adiabatic.
Adiabaticity is not a necessary consequence of inflation though and 
many inflationary models have been constructed where
isocurvature perturbations would have generically been concomitantly
produced (see e.g. \cite{langlois}, \cite{gordon}, \cite{bartolo}).
Pure isocurvature perturbations are highly excluded by present CMB
data (\cite{peiris}). Due to degeneracies with other cosmological
parameters, the amount of isocurvature modes is still weakly
constrained by the present data (see e.g. \cite{peiris},
 \cite{jussi}, \cite{crotty}).

\item Modified recombination.

The standard recombination process can be modified in several
ways. Extra sources of ionizing and resonance radiation at 
recombination or having a time-varying fine-structure constant, 
for example, can delay the recombination process and leave an
imprint on the CMB anisotropies. The present data is 
in agreement with the standard recombinations scheme. However,
non-standard recombination scenarios are still consistent with the 
current data and  may affect the current WMAP constraints on inflationary 
parameters like the spectral index, $n_{s}$, and its running
(see e.g. \cite{avelino}, \cite{martins}, \cite{bms}).

\item Neutrino Physics

The effective number of neutrinos and their effective 
mass can be both constrained by combining cosmological data.
The combination of present cosmological data under the assumption 
of several priors provide a constraint on the effective
neutrino mass of $m_{ee} <0.23 eV$ (\cite{spergel}). 
The data constraints the effective number of neutrino
species to about $N_{eff} < 7$ 
(see e.g. \cite{pierpaoli}, \cite{julien},\cite{hannestad}).

\item Dark Energy and its equation of state

The discovery that the universe's evolution may be dominated by
an effective cosmological constant is one of the most remarkable 
cosmological findings of recent years.
Observationally distinguishing a time variation in
the equation of state or finding it different from $-1$ is a 
powerful test for the cosmological constant.
The present constraints on $w$ obtained
combining the CMB data with several other cosmological 
datasets are consistent with $w=-1$, with models with
$w <-1$ slightly preferred (see e.g. \cite{mmot}, 
\cite{wlewis}). Other dark energy parameters, like
its sound speed, are weakly constrained (\cite{bo}, 
\cite{caldwell}).

\end{itemize}

\section{Conclusions}

The recent CMB data represent a beautiful success for the 
standard cosmological model. The acoustic oscillations in
the CMB temperature power spectrum, a major
prediction of the model, have now been detected 
with high statistical significance. The amplitude and
shape of the cross correlation temperature-polarization 
power spectrum is also in agreement with the expectations.

Furthermore, when constraints on cosmological parameters are 
derived under the assumption of adiabatic primordial perturbations
their values are in agreement with the predictions of the theory
and/or with independent observations.

The largest discrepancy between the standard predictions and
the data seems to come from the low value of the CMB quadrupole.
New physics has been proposed to explain this discrepancy
(see e.g. \cite{contaldi}, \cite{freese}, \cite{luminet}),
but the statistical significance is of difficult 
interpretation.

As we saw in the previous section modifications 
as extra background of relativistic particles, 
isocurvature modes or non-standard recombination schemes are still
compatible with current CMB observations, but are not necessary and
can be reasonably constrained when complementary datasets are included.\\

{\bf Acknowledgements}

I wish to thank the organizers of the conference: 
Norma Mankoc Borstnik and  Holger Bech Nielsen.
Many thanks also to Rachel Bean, Will Kinney, 
Rocky Kolb, Carlos Martins, Laura Mersini, Roya Mohayaee, 
Carolina Oedman, Antonio Riotto, Graca Rocha, Joe Silk, and Mark Trodden 
for comments, discussions and help.


\begin{thebibliography}{99}

\bibitem{afshordi}
N.~Afshordi, Y.~S.~Loh and M.~A.~Strauss,
arXiv:astro-ph/0308260.




\bibitem{ACFM}A. Albrecht,  D. Coulson, P.G. Ferreira and
	J. Magueijo, {\em Phys. Rev. Lett.} {\bf 76}, 1413 (1996).

\bibitem{avelino}
P.~P.~Avelino {\it et al.},
Phys.\ Rev.\ D {\bf 64} (2001) 103505
[arXiv:astro-ph/0102144].


\bibitem{barger}
V.~Barger, H.~S.~Lee and D.~Marfatia,
Phys.\ Lett.\ B {\bf 565} (2003) 33
[arXiv:hep-ph/0302150].

\bibitem{bartolo} N.~Bartolo, S.~Matarrese and A.~Riotto,
Phys.\ Rev.\ D {\bf 64} (2001) 123504
[arXiv:astro-ph/0107502].

\bibitem{freese}
M.~Bastero-Gil, K.~Freese and L.~Mersini-Houghton,
index,''
arXiv:hep-ph/0306289.


\bibitem{bms}
R.~Bean, A.~Melchiorri and J.~Silk,
Phys.\ Rev.\ D {\bf 68} (2003) 083501
[arXiv:astro-ph/0306357].

\bibitem{bo}
R.~Bean and O.~Dore,
arXiv:astro-ph/0307100.


\bibitem{bennett}
C.~L.~Bennett {\it et al.},
Astrophys.\ J.\ Suppl.\  {\bf 148} (2003) 1
[arXiv:astro-ph/0302207].

\bibitem{benoit}
A.~Benoit {\it et al.}  [Archeops Collaboration],
Astron.\ Astrophys.\  {\bf 399} (2003) L19
[arXiv:astro-ph/0210305].

\bibitem{review4} J.~R.~Bond,
Class.\ Quant.\ Grav.\  {\bf 15} (1998) 2573.

\bibitem{boughn}
S.~Boughn and R.~Crittenden,
arXiv:astro-ph/0305001.


\bibitem{bridle}
S.~L.~Bridle, A.~M.~Lewis, J.~Weller and G.~Efstathiou,
Mon.\ Not.\ Roy.\ Astron.\ Soc.\  {\bf 342} (2003) L72
[arXiv:astro-ph/0302306].


\bibitem{burles}
S.~Burles, K.~M.~Nollett and M.~S.~Turner,
Astrophys.\ J.\  {\bf 552}, L1 (2001)
[arXiv:astro-ph/0010171].

\bibitem{caldwell}
R.~R.~Caldwell and M.~Doran,
arXiv:astro-ph/0305334.


\bibitem{cen}
R.~Cen,
Astrophys.\ J.\  {\bf 591} (2003) L5
[arXiv:astro-ph/0303236].


\bibitem{ciardi}
B.~Ciardi, A.~Ferrara and S.~D.~M.~White,
Mon.\ Not.\ Roy.\ Astron.\ Soc.\  {\bf 344} (2003) L7
[arXiv:astro-ph/0302451].

\bibitem{contaldi}
C.~R.~Contaldi, M.~Peloso, L.~Kofman and A.~Linde,
JCAP {\bf 0307} (2003) 002
[arXiv:astro-ph/0303636].

\bibitem{crotty}
P.~Crotty, J.~Garcia-Bellido, J.~Lesgourgues and A.~Riazuelo,
Phys.\ Rev.\ Lett.\  {\bf 91} (2003) 171301
[arXiv:astro-ph/0306286].

\bibitem{julien}
P.~Crotty, J.~Lesgourgues and S.~Pastor,
Phys.\ Rev.\ D {\bf 67} (2003) 123005
[arXiv:astro-ph/0302337].


\bibitem{cyburt}
R.~H.~Cyburt, B.~D.~Fields and K.~A.~Olive,
Phys.\ Lett.\ B {\bf 567} (2003) 227
[arXiv:astro-ph/0302431].


\bibitem{copeland} E.~J.~Copeland, I.~J.~Grivell and A.~R.~Liddle,
arXiv:astro-ph/9712028.

\bibitem{dode}
S.~Dodelson,
AIP Conf.\ Proc.\  {\bf 689} (2003) 184
[arXiv:hep-ph/0309057].

\bibitem{dore}
O.~Dore, G.~P.~Holder and A.~Loeb,
arXiv:astro-ph/0309281.

\bibitem{knox} S.~Dodelson and L.~Knox,
Phys.\ Rev.\ Lett.\  {\bf 84}, 3523 (2000)
[arXiv:astro-ph/9909454].

\bibitem{doste}S.~Dodelson and E.~Stewart,
arXiv:astro-ph/0109354.


\bibitem{review5} R.~Durrer, arXiv:astro-ph/0109522.


\bibitem{efstathiou2}
G.~Efstathiou,
arXiv:astro-ph/0310207.

\bibitem{freedman} W. Freedman {\it et al.}, 
Astrophysical Journal, 553, 2001, 47.

\bibitem{gordon} C.~Gordon, D.~Wands, B.~A.~Bassett and R.~Maartens,
Phys.\ Rev.\ D {\bf 63} (2001) 023506
[arXiv:astro-ph/0009131].


\bibitem{grainge}
K.~Grainge {\it et al.},
Mon.\ Not.\ Roy.\ Astron.\ Soc.\  {\bf 341} (2003) L23
[arXiv:astro-ph/0212495].

\bibitem{kkmr}
W.~H.~Kinney, E.~W.~Kolb, A.~Melchiorri and A.~Riotto,
arXiv:hep-ph/0305130.

\bibitem{kogut}
A.~Kogut {\it et al.},
Astrophys.\ J.\ Suppl.\  {\bf 148} (2003) 161
[arXiv:astro-ph/0302213].

\bibitem{komatsu}
E.~Komatsu {\it et al.},
Astrophys.\ J.\ Suppl.\  {\bf 148} (2003) 119
[arXiv:astro-ph/0302223].

\bibitem{koso2}
A.~Kosowsky,
arXiv:astro-ph/9811163.



\bibitem{kosowsky} A.~Kosowsky and M.~S.~Turner,
Phys.\ Rev.\ D {\bf 52} (1995) 1739
[arXiv:astro-ph/9504071].


\bibitem{halverson} N.~W.~Halverson {\it et al.},
arXiv:astro-ph/0104489.

\bibitem{hannestad}
S.~Hannestad,
JCAP {\bf 0305} (2003) 004
[arXiv:astro-ph/0303076].


\bibitem{review2} W.~Hu, D.~Scott, N.~Sugiyama and M.~J.~White,
Phys.\ Rev.\ D {\bf 52}, 5498 (1995)
[arXiv:astro-ph/9505043].

\bibitem{review} W.~Hu, N.~Sugiyama and J.~Silk,
Nature {\bf 386}, 37 (1997)
[arXiv:astro-ph/9604166].

\bibitem{dasipol}
J.~Kovac, E.~M.~Leitch, P.~C., J.~E.~Carlstrom, H.~N.~W. and W.~L.~Holzapfel,
Nature {\bf 420} (2002) 772
[arXiv:astro-ph/0209478].

\bibitem{acbar}
C.~l.~Kuo {\it et al.}  [ACBAR collaboration],
arXiv:astro-ph/0212289.

\bibitem{langlois} D.~Langlois and A.~Riazuelo,
Phys.\ Rev.\ D {\bf 62} (2000) 043504.


\bibitem{leach}
S.~M.~Leach and A.~R.~Liddle,
arXiv:astro-ph/0306305.


\bibitem{lee} A.~T.~Lee {\it et al.},
Astrophys.\ J.\  {\bf 561} (2001) L1
[arXiv:astro-ph/0104459].

\bibitem{luminet}
J.~P.~Luminet, J.~Weeks, A.~Riazuelo, R.~Lehoucq and J.~P.~Uzan,
Nature {\bf 425} (2003) 593
[arXiv:astro-ph/0310253].


\bibitem{martins}
C.~J.~A.~Martins, A.~Melchiorri, G.~Rocha, R.~Trotta, P.~P.~Avelino and P.~Viana,
arXiv:astro-ph/0302295.


\bibitem{mauskopf} P.~D.~Mauskopf {\it et al.}  [Boomerang Collaboration],
Astrophys.\ J.\  {\bf 536}, L59 (2000)
[arXiv:astro-ph/9911444].


\bibitem{mmot}
A.~Melchiorri, L.~Mersini, C.~J.~Odman and M.~Trodden,
Phys.\ Rev.\ D {\bf 68} (2003) 043509
[arXiv:astro-ph/0211522].

\bibitem{odman}
A.~Melchiorri and C.~Odman,
Phys.\ Rev.\ D {\bf 67} (2003) 081302
[arXiv:astro-ph/0302361].

\bibitem{melksilk} A.~Melchiorri and J.~Silk,
arXiv:astro-ph/0203200.

\bibitem{melchiorri} A.~Melchiorri {\it et al.}  [Boomerang Collaboration],
Astrophys.\ J.\  {\bf 536} (2000) L63
[arXiv:astro-ph/9911445].

\bibitem{miller} A.~D.~Miller {\it et al.},
Astrophys.\ J.\  {\bf 524}, L1 (1999)
[arXiv:astro-ph/9906421].

\bibitem{netterfield} C.~B.~Netterfield {\it et al.}  
[Boomerang Collaboration], arXiv:astro-ph/0104460.


\bibitem{nolta}
M.~R.~Nolta {\it et al.},
arXiv:astro-ph/0305097.


\bibitem{costa}
A.~de Oliveira-Costa, M.~Tegmark, M.~Zaldarriaga and A.~Hamilton,
arXiv:astro-ph/0307282.



\bibitem{page}
L.~Page {\it et al.},
arXiv:astro-ph/0302220.


\bibitem{pearson}
T.~J.~Pearson {\it et al.},
Astrophys.\ J.\  {\bf 591} (2003) 556
[arXiv:astro-ph/0205388].

\bibitem{Peeb1970} P.J.E. Peebles, and Yu, J.T. 1970,
Ap.J. 162, 815  


\bibitem{peiris}
H.~V.~Peiris {\it et al.},
Astrophys.\ J.\ Suppl.\  {\bf 148} (2003) 213
[arXiv:astro-ph/0302225].

\bibitem{pierpaoli}
E.~Pierpaoli,
Mon.\ Not.\ Roy.\ Astron.\ Soc.\  {\bf 342} (2003) L63
[arXiv:astro-ph/0302465].


\bibitem{scranton}
R.~Scranton {\it et al.}  [SDSS Collaboration],
arXiv:astro-ph/0307335.


\bibitem{anze}
A.~Slosar {\it et al.},
Mon.\ Not.\ Roy.\ Astron.\ Soc.\  {\bf 341} (2003) L29
[arXiv:astro-ph/0212497].

\bibitem{spergel}
D.~N.~Spergel {\it et al.},
Astrophys.\ J.\ Suppl.\  {\bf 148} (2003) 175
[arXiv:astro-ph/0302209].



\bibitem{SZ70}Sunyaev, R.A. \& Zeldovich,
Ya.B., 1970,  Astrophysics and Space Science 7, 3 


\bibitem{tegb97} M.~Tegmark,
Astrophys.\ J.\  {\bf 514}, L69 (1999)
[arXiv:astro-ph/9809201].


\bibitem{torbet} E.~Torbet {\it et al.},
Astrophys.\ J.\  {\bf 521}, L79 (1999)
[arXiv:astro-ph/9905100].


\bibitem{jussi}
J.~Valiviita and V.~Muhonen,
Phys.\ Rev.\ Lett.\  {\bf 91} (2003) 131302
[arXiv:astro-ph/0304175].

\bibitem{verde}
L.~Verde {\it et al.},
Astrophys.\ J.\ Suppl.\  {\bf 148} (2003) 195
[arXiv:astro-ph/0302218].


\bibitem{wang2}
X.~Wang, M.~Tegmark, B.~Jain and M.~Zaldarriaga,
arXiv:astro-ph/0212417.

\bibitem{wlewis}
J.~Weller and A.~M.~Lewis,
arXiv:astro-ph/0307104.

\bibitem{review3} M.~J.~White, D.~Scott and J.~Silk,
Ann.\ Rev.\ Astron.\ Astrophys.\  {\bf 32} (1994) 319.

\bibitem{wilson} M.~L.~Wilson and J.~Silk,
Astrophys.\ J.\  {\bf 243} (1981) 14.

\end{thebibliography}
\end{document}